\documentclass[prl,showpacs,twocolumn]{revtex4}
\usepackage{mathrsfs}
\usepackage{amsmath}
\usepackage{amssymb}
\usepackage{revsymb}
\usepackage{graphicx}
\usepackage{mathrsfs}
\begin{document}
\title{Non-perturbative vacuum-polarization effects in proton-laser collisions}
\author{A. \surname{Di Piazza}}
\email{dipiazza@mpi-hd.mpg.de}
\author{K. Z. \surname{Hatsagortsyan}}
\email{k.hatsagortsyan@mpi-hd.mpg.de}
\author{C. H. \surname{Keitel}}
\email{keitel@mpi-hd.mpg.de}
\affiliation{Max-Planck-Institut f\"ur Kernphysik, Saupfercheckweg 1, D-69117 Heidelberg, Germany}

\date{\today}

\begin{abstract}
In the collision of a high-energy proton beam and a strong laser field, merging of the laser photons can occur due to the polarization of vacuum. The probability of photon merging is calculated by accounting exactly for the laser field and presents a highly non-perturbative dependence on the laser intensity and frequency. It is shown that the non-perturbative vacuum-polarization effects can be experimentally measured by combining the next-generation of table-top petawatt lasers with presently available proton accelerators.

 
\end{abstract}
 
\maketitle

In classical electrodynamics the superposition principle holds for electromagnetic fields of arbitrary strength in vacuum. Instead, quantum electrodynamics (QED) predicts that under the influence of electromagnetic fields of the order of the ``critical'' fields: $E_{cr}=m^2c^3/\hbar e=1.3\times 10^{16}\;\text{V/cm}$ and $B_{cr}=m^2c^3/\hbar e=4.4\times 10^{13}\;\text{G}$, the vacuum behaves as a nonlinear medium ($-e<0$ and $m$ are the electron charge and mass, respectively) \cite{Dittrich_b_2000}. The experimental observation of nonlinear vacuum-polarization effects (VPEs) is of fundamental interest, especially to test the validity of QED at fields of such high strength. Highly charged ions with a charge number $Z$ such that $Z\alpha\lesssim 1$ with $\alpha=e^2/4\pi$ being the fine-structure constant, provide electric fields of the order of $E_{cr}$ at the typical QED length scale $\lambda_c=\hbar/mc$ (Compton length).  This has allowed the observation of VPEs induced by the strong Coulomb field of heavy ions via the measurement of Delbr\"{u}ck scattering, i. e. the scattering of a $\gamma$-photon by an atomic field \cite{DS} and of the related process of $\gamma$-photon splitting in atomic fields \cite{PSA}. Noticeably, in both cases effects of high-order corrections in $Z\alpha$ have been observed in the cross-section \cite{Milstein}.

The continuous progress in laser technology opens unique opportunities to test QED in the presence of intense electromagnetic waves \cite{Review}. The record laser intensity of $7\times 10^{21}\;\text{W/cm$^2$}$ was produced in laboratory by focusing a 45 TW Ti:Sapphire laser onto one wavelength \cite{Bahk_2004}. Moreover, the Optical Parametric Chirped Pulse Amplification (OPCPA) technique has already been successfully applied to generate terawatt pulses with a short duration ($10\;\text{fs}$) at a high repetition rate ($30\;\text{Hz}$) \cite{Witte_2005}. A petawatt laser is currently under development based on the same technique \cite{PFS}. Finally, envisaged intensities of the order of $10^{24}\text{-}10^{26}\;\text{W/cm$^2$}$ are likely to be reached in the near future at the Extreme Light Infrastructure (ELI) \cite{ELI}. The first theoretical investigations of QED in the presence of a strong background laser field go back to the sixties \cite{Reiss_1962,Nikishov_1964}. Later, the calculations of higher-order QED diagrams like the mass operator \cite{QEDLaser}, the polarization operator \cite{QEDLaser2} and the photon splitting \cite{Di_Piazza_2007} were performed. Refractive VPEs induced by laser fields have never been experimentally observed, essentially because present laser intensities are much smaller than the critical intensity $I_{cr}=E_{cr}^2/8\pi=2.3\times 10^{29}\;\text{W/cm$^2$}$ and laser photon energies much smaller than $mc^2=0.5\;\text{MeV}$. To the best of our knowledge, the only feasible proposals to observe refractive VPEs in a laser field explored so far in the literature are restricted to perturbative VPEs in which the leading contributions in the laser field are the only, at least theoretically, observable effects while all higher-order effects are strongly suppressed \cite{VPEs}.

In this Letter we put forward a scheme for which it is experimentally feasible to observe non-perturbative refractive VPEs in laser fields. We calculate the probability for an even number of laser photons to merge in the head-on collision of a high-energy proton beam and a strong laser field due to VPEs, by accounting exactly for the laser field and by focusing on the non-perturbative dependence of this probability on the laser intensity $I_0$ and frequency $\omega_0$. As a result, non-perturbative QED can be tested at the opening of multiphoton vacuum polarization channels.

To this end the proton beam has unique features. On the one hand, the proton is light enough to be accelerated up to very high energies, namely to $980\;\text{GeV}$ at the Tevatron, or even to $7\;\text{TeV}$ at the Large Hadron Collider (LHC) \cite{PDG}. Then, the laser field amplitude and frequency in the proton rest frame are enhanced by the Lorentz $\gamma$-factor of the proton so that the nonlinear QED regime becomes reachable, allowing in principle for the experimental observation of high-order non-perturbative VPEs. On the other hand, the proton mass $M=938\;\text{MeV}$ is large enough that the background radiation arising from multiphoton Thomson scattering of the laser photons is, as we will see, very low.  Moreover, since the proton charge number is $Z=1$, here the VPEs are essentially driven by the strong laser field. 

Without loss of generality we can consider a proton moving along the negative $y$ direction with velocity $\beta$ and a counterpropagating laser beam with amplitude $E_0=\sqrt{8\pi I_0}$ and frequency $\omega_0$ (from now on natural units with $\hbar=c=1$ are used). The laser is assumed to be linearly polarized along the $z$ direction.  The process of laser photons merging due to VPEs in a proton field is represented by the Feynman diagram in Fig. 1
\begin{figure}
\begin{center}
\includegraphics[width=7cm]{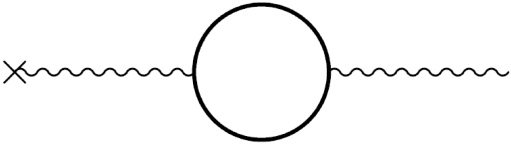}
\end{center}
\caption{Feynman diagram corresponding to the process of laser photon merging induced by VPEs in a proton field.}
\end{figure}
where the right photon leg represents the outgoing photon resulting from the merging, the thick fermion loop indicates that the fermion propagators are calculated by accounting exactly for the laser field and where the crossed photon leg indicates the electromagnetic field of the proton taken into account as a perturbation since $Z=1$ and described by the four-potential
$A^{\mu}(t,\mathbf{r})=eu^{\mu}/4\pi\sqrt{x^2+\gamma^2(y+\beta t)^2+z^2}$
with $u^{\mu}=(\gamma,0,-\beta\gamma,0)$ being the proton four-velocity and $\gamma=1/\sqrt{1-\beta^2}$ its relativistic Lorentz factor. The probability amplitude of the laser-photon merging process can be calculated as $T(k)=(2\pi)^{-4}\int d^4q A_{\mu}(q)T^{\mu\nu}(q,k)e^*_{\nu}(k)/\sqrt{2\omega}$ (see Fig. 1)
where $A_{\mu}(q)$ is the four-dimensional Fourier transform of $A_{\mu}(t,\mathbf{r})$, $T^{\mu\nu}(q,k)$ is the polarization tensor in a laser field  \cite{QEDLaser2} and $e_{\nu}(k)$ is the polarization four-vector of the outgoing real photon with four-momentum $k^{\mu}=(\omega,\mathbf{k})$ \cite{Landau_b_4_1982}. Starting from the above amplitude, one obtains the differential rate $d\mathcal{R}$ by applying Fermi's golden rule: $d\mathcal{R}=|T(k)|^2d\mathbf{k}/(2\pi)^3$ (the quantization space-volume and time-interval are set equal to unity here) \cite{Landau_b_4_1982}. In the considered case of a monochromatic plane wave the total rate can be written as $d\mathcal{R}=\sum_{n=1}^{\infty}d\mathcal{R}_n$ where $d\mathcal{R}_n$ is the rate of photons resulting from the merging of $2n$ laser photons (Furry's theorem prevents the merging of an odd number of laser photons in this process \cite{Landau_b_4_1982}). The integration over $\mathbf{k}$ is easier by employing spherical coordinates $(\omega,\vartheta,\varphi)$ with the $y$-axis as polar axis. The integrals over $\omega$ and $\varphi$ can be carried out and the ensuing differential rate $d\mathcal{R}_n/d\vartheta$ written as
\begin{equation}
\label{dRdtheta_n}
\frac{d\mathcal{R}_n}{d\vartheta}=\frac{\alpha^3}{64\pi^2}\frac{(1+\beta)m^4}{\omega_0^3}\frac{\sin^3\vartheta}{(1-\cos\vartheta)^4}\frac{|c_{1,n}|^2+|c_{2,n}|^2}{n^3}.
\end{equation}
In this expression $\vartheta$ is the angle between the momentum of the emitted photon and the $y$-axis and
\begin{widetext}
\begin{equation}
\label{c_j_n_ex}
c_{j,n}=i^n\int_{-1}^1dv\int_0^{\infty}\frac{d\rho}{\rho}e^{-i\Phi_n}\left\{\xi^2\left[\frac{A}{2}\mathcal{J}^*_n(z_n)-\sin^2\rho J_n(z_n)\right]\delta_{j,1}+\xi^2\frac{\sin^2\rho}{1-v^2}\mathcal{J}_n(z_n)+\frac{\eta_n}{2}\left(n-i\frac{1-v^2}{4\rho}\right)J_n(z_n)\right\}
\end{equation}
\end{widetext}
where $j\in\{1,2\}$, $\delta_{j,j'}$ is the Kronecker $\delta$-function, $\Phi_n=2n\rho+4\rho\{1+\xi^2[1-\sin^2(\rho)/\rho^2]/2\}/\eta_n(1-v^2)$, $z_n=2\rho\xi^2[\sin^2(\rho)/\rho^2-\sin (2\rho)/2\rho]/\eta_n(1-v^2)$, $A=1+\sin^2(\rho)/\rho^2-\sin (2\rho)/\rho$ and $\mathcal{J}_n(z)=J_n(z)+iJ'_n(z)$ with $J_n(z)$ being the ordinary Bessel function of order $n$ and $J'_n(z)$ its derivative. The coefficients $c_{j,n}$ depend on the two Lorentz- and gauge-invariant quantities $\xi=eE_0/m\omega_0$ and $\eta_n=(k_0k_n)/m^2=\omega_0\omega_n(1-\cos\vartheta)/m^2$ where $k_0^{\mu}=(\omega_0,0,\omega_0,0)$ is the four-momentum of the laser photons and $k_n^{\mu}$ is the four-momentum of the outgoing photon whose energy $\omega_n=2n\omega_0(1+\beta)/(1+\beta\cos\vartheta)$ depends on the number of laser photons merged, according to the energy conservation and the Doppler-shift. 
When the proton is highly relativistic, $\omega_n$ is 4$\gamma^2$ times larger than $2n\omega_0$ for backscattered photons.

In the asymptotic limit $\xi \ll 1$ of weak laser fields, the leading rate in Eq. (\ref{dRdtheta_n}) is that with $n=1$ which coincides with the result of the Feynman box diagram corresponding to the perturbative treatment of the laser field. The rate increases with raising $\xi$. Moreover, present optical lasers easily exceed the relativistic threshold $\xi\approx 1$ (laser intensity $I_0\approx 10^{18}\;\text{W/cm$^2$}$ at $\omega_0\approx 1\;\text{eV}$). Therefore, we will discuss here only the most practical regime $\xi\gg 1$. 
In this asymptotic limit the dynamics is determined by the parameter $\chi_n$:
\begin{equation}
\label{chi_n}
\chi_n\equiv\xi\eta_n=\sqrt{\frac{I_0}{I_{cr}}}\frac{2n(1+\beta)\omega_0}{m}\frac{1-\cos\vartheta}{1+\beta\cos\vartheta}.
\end{equation}
The leading-order contribution to the coefficients $c_{j,n}$ depends only on $\chi_n$ and is given by
\begin{equation}
c_{j,n}=e^{-i\pi/3}\int_0^1dv\int_0^{\infty}\frac{d\lambda}{\lambda}e^{-\exp(i\pi/3)\lambda-x_n}b_{j,n}
\end{equation}
where $x_n=\chi_n^2\lambda^3(1-v^2)^2/96$ and
\begin{equation}
b_{j,n}=j\chi_n^2\lambda^2\frac{1-v^{4/j}}{16}[I_n(x_n)-I'_n(x_n)]+\frac{I_n(x_n)}{\lambda}
\end{equation}
with $I_n(x)$ being the modified Bessel function of order $n$ and $I'_n(x)$ its derivative. As the above expressions show, the dependence of the differential rate $d\mathcal{R}_n/d\vartheta$ on the parameter $\chi_n$
is highly non-perturbative. The perturbative result is obtained only if $\chi_n\ll 1$ (for the process at hand the leading contributions at $\xi\gg 1$ and $\chi_n\ll 1$ coincide with those at $\xi\ll 1$ and $\eta_n\ll 1$). In this case, the leading order asymptotics of the coefficients $c_{j,n}$ have the typical perturbative scaling as $\chi_n^{2n}$:
\begin{equation}
\label{c_n_pert}
c_{j,n}=(-1)^{n+1}\sqrt{\pi}\frac{(3n-2)!}{\Gamma(2n+3/2)}\frac{(2n)!}{n!}\frac{3jn+1}{192^n}\chi_n^{2n}
\end{equation}
with Gamma function $\Gamma(x)$. In the particular case $n=1$ and $\beta=0$ our expression of the total rate $\mathcal{R}_1=\int_0^{\pi} d\vartheta d\mathcal{R}_1/d\vartheta$ is in agreement with the results in the paper by Milstein \textit{et al.} in \cite{VPEs}.

In the opposite limit of large $\chi_n$ we obtain
\begin{equation}
c_{j,n}=\frac{2e^{2i\pi/3}}{21}\frac{j+1}{6^{1/3}}\frac{\Gamma(2/3)\Gamma(5/6)}{\Gamma(7/6)}\frac{\Gamma(n-1/3)}{\Gamma(n+4/3)}\chi_n^{2/3}.
\end{equation}
In this regime of ultra strong laser fields where $\chi_n \gg 1$, the differential rate scales non-perturbatively with $\chi_n$, the higher corrections in $\chi_n$ stemming from exchanges of laser photons with no \emph{net} absorption. However, the rate decreases with $n$ monotonically and behaves at large $n$ as $d\mathcal{R}_n/d\vartheta\sim 1/n^5$. 

We consider below some numerical examples corresponding to different possible experimental conditions in which the process of photon merging is feasible. In the first one we consider the interaction between a petawatt laser pulse and a proton bunch with parameters available at the LHC. Concerning the laser, we use the following data \cite{PFS,ELI}: pulse energy $3\;\text{J}$, pulse duration $5\;\text{fs}$ at $10\;\text{Hz}$ repetition rate. When the beam is focused onto one wavelength of $0.8\;\text{$\mu$m}$ its intensity is $I_0=3\times 10^{22}\;\text{W/cm$^2$}$. The main parameters of the proton bunch are \cite{PDG}: proton energy $7\;\text{TeV}$, number of protons per bunch $11.5\times 10^{10}$, bunch transversal radius $16.6\;\text{$\mu$m}$, bunch length $7.55\;\text{cm}$. In Fig. 2 we show the differential rate $d\mathcal{R}_1/d\vartheta$ of the photons resulting from the merging of two laser photons (continuous line). In the upper horizontal axis in Fig. 2 we have indicated the value of the parameter $\chi_1$ at the corresponding angle $\vartheta$: it is evident that in the relevant region of the spectrum $\chi_1\sim 1$ and the perturbative approach is inadequate. For instance, the perturbative result for the rate $d\mathcal{R}_1/d\vartheta$ at $\vartheta=3.1415$ is roughly $28\;\%$ less than the non-perturbative one. 
\begin{figure}
\begin{center}
\includegraphics[width=8cm]{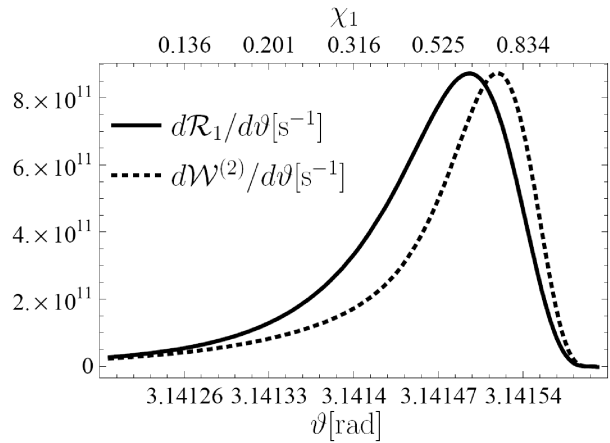}
\end{center}
\caption{The rate per unit angle $\vartheta$ of photons emitted via VPEs (continuous line) and via two-photon Thomson scattering (dashed line). The upper horizontal axis shows the values of the parameter $\chi_1$ as a function of $\vartheta$ [see Eq. (\ref{chi_n})].}
\end{figure}
In this regime there is a competing process: the two-photon Thomson scattering of the laser photons by the proton beam. In fact, the energies of the photons produced via the photon merging and via the two-photon Thomson scattering are the same because of the negligible recoil effect in the latter case (the photon energy is $\approx 470\;\text{MeV}$ at the maximal emission angle, while the proton energy is $7\;\text{TeV}$). In Fig. 2 the vacuum polarization signal is compared with the analogous quantity $d\mathcal{W}^{(2)}/d\vartheta$ of the two-photon Thomson scattering of the laser photons by the proton beam (dashed line, result via \cite{Nikishov_1964}). The figure shows that the vacuum polarization contribution is clearly distinguishable from the two-photon Thomson scattering. By integrating $d\mathcal{R}_1/d\vartheta$ with respect to $\vartheta$ and by multiplying the ensuing quantity with laser pulse duration and repetition rate and the effective number of protons that interact with the laser beam, we obtain approximately 400 events per hour. Moreover, due to the non-perturbative nature of the process in this regime, the rate of photons resulting from the merging of \emph{four} laser photons is not small either: 5.4 events per hour. As expected, in this case the four-photon Thomson scattering is safely negligible due to its scaling as $\xi_p^8$, with $\xi_p=(m/M)\xi\approx 6.4\times 10^{-2}$. In the previous proposals \cite{VPEs}, the corresponding multiphoton channels involving more laser photons than the minimum required, are suppressed by several orders of magnitude. 

In the following example we consider the possibility of employing the already operative Tevatron accelerator. The relevant parameters of the proton bunches at the Tevatron are \cite{PDG}: proton energy $980\;\text{GeV}$, number of protons per bunch $24\times 10^{10}$, bunch transversal radius $29\;\text{$\mu$m}$, bunch length $50\;\text{cm}$. The relatively small proton energy in this case can be compensated by employing strong attosecond pulses of XUV radiation which are proposed in \cite{Tsakiris_2006}. In particular, we use an attosecond pulse of intensity $I_0=3.8\times 10^{22}\;\text{W/cm$^2$}$ with a photon energy of $70\;\text{eV}$ and a pulse duration of $60\;\text{as}$. The latter, according to the simulations in \cite{Tsakiris_2006}, can be produced by the reflection from a plasma surface of a 6 PW Ti:Sapphire laser pulse of $5\;\text{fs}$ duration focused onto a spot-radius of $5\;\text{$\mu$m}$. In this setup, the angular distribution of the photons emitted via VPEs has a shape similar to that in Fig. 2 and the photon rate produced via two-photon Thomson scattering is negligible. We report only the average number of photons emitted in one hour by the merging of two and of four laser photons, which is approximately equal to 490 and 6.6, respectively. As in \cite{Tsakiris_2006} we have assumed that about $6\times 10^{-2}$ of the energy of the initial beam is transferred to the attosecond pulse and that the OPCPA technique is employed to achieve a repetition rate of $10\;\text{Hz}$. Also in this case the values of the parameter $\chi_n$ near the maximum of the photon angular spectrum does not permit the use of perturbation theory as $\chi_1=\chi_2/2\approx 0.7$ at $\vartheta=3.1410$. At the same angle the energy of the emitted photon is about $440\;\text{MeV}$. We stress that in this example the relativistic parameter $\xi$ of the XUV pulse is approximately equal to three, which is not much larger than unity as it is assumed in our model. In this case, high-order corrections are expected to be about $1/\xi$ or $\eta_n$ times, i. e.   $\approx 30\text{-}40\;\%$ of, our leading-order result. For the benefit of a quantitative analysis, the exact expressions (\ref{c_j_n_ex}) of the coefficients $c_{j,n}$ have to be used in Eq. (\ref{dRdtheta_n}).

By employing the laser parameters of ELI ($70\;\text{PW}$ Ti:Sapphire laser focused onto $0.8\;\text{$\mu$m}$, $5\;\text{fs}$ pulse duration at $1\;\text{Hz}$ repetition rate) we would obtain up to two photons per pulse from two-photon merging and even the merging of \emph{ten} laser photons would be experimentally accessible (two photons per hour).

In the collision of a laser pulse with a proton beam Delbr\"uck scattering can occur as well. While this process is clearly distinguishable from the process at hand insofar that the frequency of the scattered photon is different from $\omega_n$ for any $n$, the question arises if the Delbr\"uck scattering can compete with the photon merging process in terms of probability. For the considered setup, the answer is no. In fact, starting from the Born-approximated cross section of the Delbr\"uck scattering which is certainly suitable at $Z=1$ (see, e. g., \cite{Milstein}), it can be shown that the order of magnitude of the ratio between the number of photons scattered via Delbr\"uck scattering and the rate $\mathcal{R}_1$ is given by $(\alpha/\xi)^2$. This quantity is very small in the situations considered above. 

In the non-perturbative regime of laser-proton interaction when $\chi_n \sim 1$, electron-positron pair creation may occur. We show that the depletion of the laser energy due to this process is negligible. Thus, in the direct channel the pairs are created as a tunneling process when $\xi\gg 1$ \cite{Pair_Prod}. 
The order of magnitude of the pair production rate at $\xi\gg 1$ and at laser intensities close to $I_{cr}$ (in the rest frame of the proton) is roughly $\alpha^2 m$. For the data in the first (second) example considered, about $10^6\text{-}10^7$ ($10^8$) pairs are produced per pulse corresponding only to a few $\mu\text{J}$ (few tens of $\mu\text{J}$) energy loss which is negligible with respect to the pulse energy of $3\;\text{J}$ ($1.8\;\text{J}$). 

The high-energy photons produced via VPEs still travel through the laser beam almost in the opposite direction and could decay via the production of electron-positron pairs. The rate of pair production by a photon with energy $\omega_1$ in a laser field with $\xi\gg 1$ and $\chi_1\sim 1$ can be estimated as $\sim 10^{-2}\alpha m^2/\omega_1$ by employing the results in \cite{QEDLaser2}. Then, the lifetime in the laser field of a photon with an energy $\sim 400\text{-}500\;\text{MeV}$ is around $14\text{-}17\;\text{fs}$, i. e. longer than the considered laser beam durations. Therefore, the photons produced via VPEs are able to escape from the region where the strong laser field is present and can in principle be detected. 

In conclusion, we have shown that existing proton accelerators combined with next generation table-top petawatt lasers permit to measure non-perturbative refractive VPEs in a laser field. These become manifest in the opening of multiphoton channels of interaction as well as in the enhanced contribution of high-order effects. 
%
%

\end{document}